\documentclass[prl,aps,twocolumn,showpacs]{revtex4-1}

\usepackage{graphicx}
\usepackage{subfig} 
\usepackage{dcolumn}
\usepackage{multirow}
\usepackage{threeparttable} 
\usepackage{soul} 
\usepackage{verbatim} 
\usepackage{array} 

\def\beq{\nopagebreak \begin{equation}}
\def\eeq{\end{equation}}

\newcolumntype{P}[1]{>{\centering\arraybackslash}p{#1}}
 
\usepackage{amsmath}
\DeclareMathOperator{\Tr}{Tr}
\usepackage{caption}
\usepackage{braket}
\usepackage{color}
\definecolor{forestgreen}{RGB}{34, 139, 34}
\usepackage{fixmath} 

\begin{document}
\title{Phonon transport unveils the prevalent point defects in GaN }

\author{Ankita Katre$^1$}
\email{ankitamkatre@gmail.com}
\author{Jes\'{u}s Carrete$^2$}
\author{Tao Wang$^3$}
\author{Georg K. H. Madsen$^2$}
\author{Natalio Mingo$^1$}
\email{natalio.mingo@cea.fr}
\affiliation{$^1$LITEN, CEA-Grenoble, 17 rue des Martyrs, 38054 Grenoble Cedex 9, France}
\affiliation{$^2$Institut f\"ur Materialchemie, Technische Universit\"at Wien, a-1060 Vienna, Austria}
\affiliation{$^3$AMS, ICAMS, Ruhr-Universit\"at Bochum, 44801 Bochum, Germany}

\begin{abstract}

Determining the types and concentrations of vacancies present in intentionally doped GaN is a notoriously difficult and long-debated problem. 
Here we use an unconventional approach, based on thermal transport modeling, 
to determine the prevalence of vacancies in previous measurements. This allows us to provide conclusive evidence of the recent hypothesis that gallium vacancies in ammonothermally grown samples can be complexed with hydrogen. 
Our calculations for O-doped and Mg-O co-doped samples  yield a consistent picture interlinking dopant and vacancy concentration, carrier density, and thermal conductivity, in excellent agreement with experimental measurements. These results also highlight the predictive power of \emph{ab initio} phonon transport modeling, and its value for understanding and quantifying defects in semiconductors.

\end{abstract}

\maketitle

GaN has revolutionized the fields of solid state lighting and power electronics \cite{nakamura_nobel_2015,jones_review_2016,nakamura_history_2013,pimputkar_prospects_2009}. In these applications the material's performace is critically dependent on its defects. However understanding GaN's defects has turned out to be a very challenging problem \cite{lyons_computationally_2017,walle_first-principles_2004,van_de_walle_interactions_1997}. In the endeavor to clarify it, theory has greatly helped to resolve long lasting controversies and reconcile seemingly conflicting results \cite{walle_universal_2003,lyons_shallow_2012,lindsay_thermal_2012,ganchenkova_nitrogen_2006,elsner_theory_1997,buckeridge_determination_2015}. Examples are the intrinsic $n$-type character of carriers, originally attributed to nitrogen vacancies but later shown to be the result of residual hydrogen and oxygen impurities \cite{boguslawski_native_1995,freysoldt_first-principles_2014, walle_first-principles_2004, Neugebauer_PRB1994}, or the explanation of the photoluminescent signatures of Mg doped samples \cite{lyons_shedding_nodate,lyons_shallow_2012}. A presently open debate concerns the interdependence between external doping, vacancies, and carrier concentration. Based on positron annihilation experiments a relation has recently been suggested between the concentration of oxygen doping, Ga vacancies, and hydrogen associated to them \cite{tuomisto_vacancyhydrogen_2014}. We provide independent evidence supporting this hypothesis from an unusual source, namely thermal transport.

The thermal conductivity ($\kappa$) is a rich and largely unexplored source of information on defects in materials. Each defect type has a signature on thermal transport. By combining measurements and 
\emph{ab initio} thermal conductivity modeling it 
has recently become
possible to decipher the type and concentration of native defect types present in the material\cite{, Katre_PRL2017,Katre_JMCA2016, Katcho_PRB2014}, even without the use of more involved defect characterization techniques such as secondary ion mass spectrometry, positron annihilation 
or
carrier lifetime spectroscopy \cite{Cheng_SSC1973, Schultz_RMP1988, Colton_JVST1981, Gaubas_ECS2016}. In the case of doped GaN, a fundamental question is then 
whether one can elucidate the types and concentrations of vacancies 
by using the calculated thermal transport signatures of the different defects.
We provide the answer here. In what follows, we present \emph{ab initio} calculations of GaN's thermal conductivity under Mg and O doping in the presence of different vacancies, and compare them with measurements on ammonothermally grown samples from ref.~\citenum{Simon_APL2014}. The resulting picture coherently relates thermal conductivities, extrinsic and intrinsic defect concentrations, and carrier densities,  
yielding
excellent agreement between theory and experiment.

In the decade elapsed since its initial demonstration \cite{Broido_APL07}, the ability to predict 
the thermal conductivity of single crystals in a parameter-free fashion has rapidly evolved to become a recognized computational tool, implemented in several software packages \cite{ALMA_BTE,  Li_CPC2014, Tadano_JPCM2014, Togo_PRB2015, Cherna_CPC2015}. In contrast, predicting the thermal conductivity of materials with vacancies is not mainstream, and it has been achieved only recently \cite{Katcho_PRB2014, Katre_JMCA2016, Katre_PRL2017}. The theory and implementation of \emph{ab initio} scattering by defects is considerably more involved than the often used formulas for mass-defect scattering, requiring the symmetrization of the \emph{ab initio} force constants around the vacancy, and a T-matrix formulation of the scattering by isolated defects \cite{Katre_PRL2017, Katcho_PRB2014, Mingo_PRB2010}, as implemented in our program almaBTE \cite{ALMA_BTE}. This approach has so far been applied to predict the effect of vacancies and substitutional impurities on the thermal conductivity of diamond \cite{Katcho_PRB2014}, determine the dominant antisite types in ZrNiSn \cite{Katre_JMCA2016}, the role of antisites in Fe$_2$VAl \cite{Bandaru_NMTE2017}, and understand the effect of substitutional impurities in SiC \cite{Katre_PRL2017}, in all cases showing good agreement with experimental results. For the sake of space, we proceed directly to discuss our results, referring to refs.~\citenum{ALMA_BTE,Katre_PRL2017, Katcho_PRB2014, Mingo_PRB2010} for an explanation of the method, and providing more details on \emph{ab initio} calculations in the ESI.

\begin{figure}[t]
 \centering
  \includegraphics[width=8.1cm]{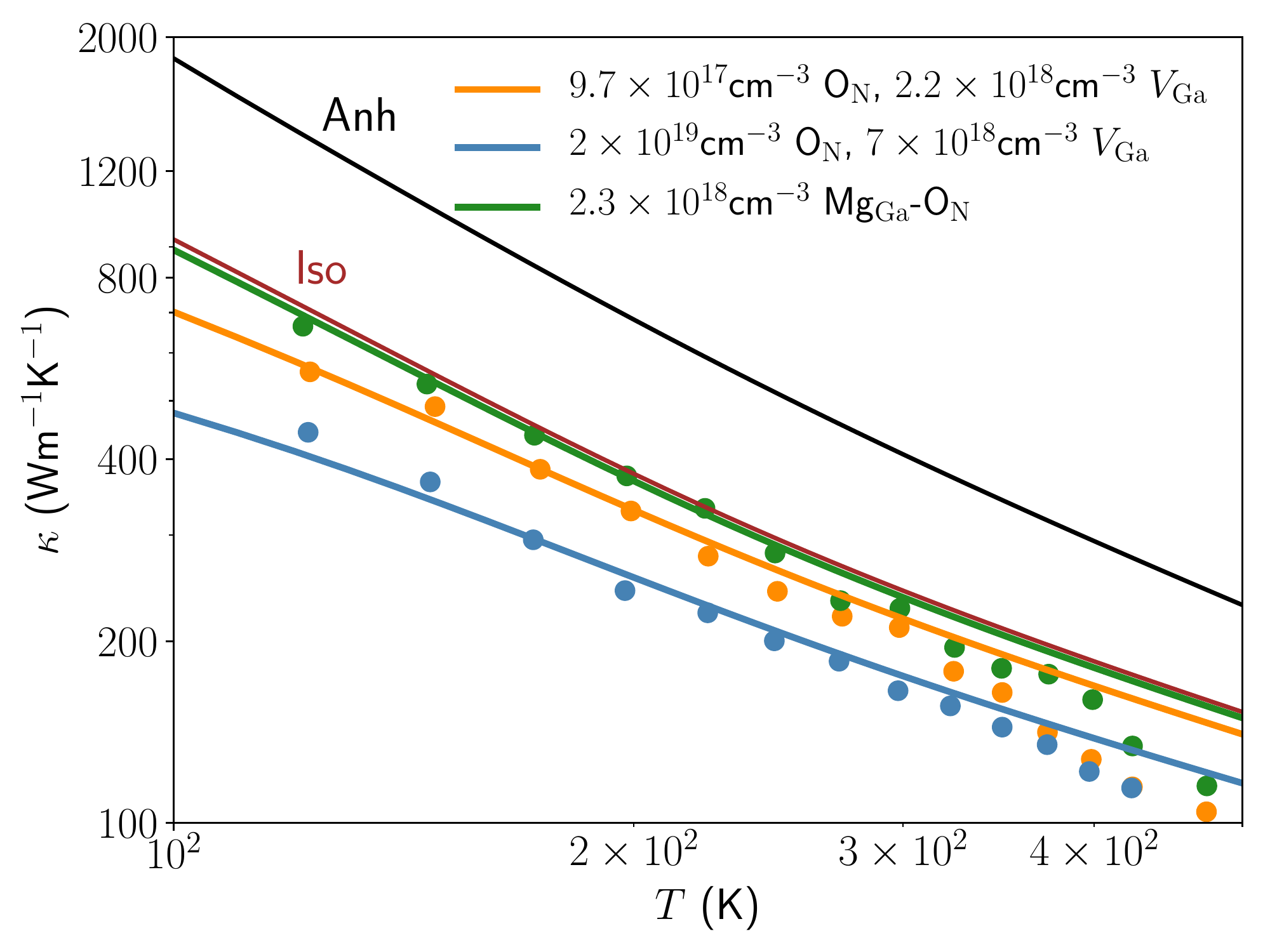} 
 \caption{Lattice thermal conductivity of GaN with O$_{\mathrm{N}}$, Mg$_{\mathrm{Ga}}$-O$_{\mathrm{N}}$ defects. O-doped samples also contain gallium vacancies ($V_{\mathrm{Ga}}$), which are included in our calculations. The vacancy concentrations of 
the respective samples are obtained from the positron annihilation experiments in ref.~\citenum{tuomisto_vacancyhydrogen_2014}. The calculated thermal conductivity results with different concentrations of defects are in good agreement with corresponding measurements from  ref.~\citenum{Simon_APL2014}.}
 \label{fig:kappaT} 
\end{figure}

\begin{figure}[t]
 \centering
  \includegraphics[width=8.3cm]{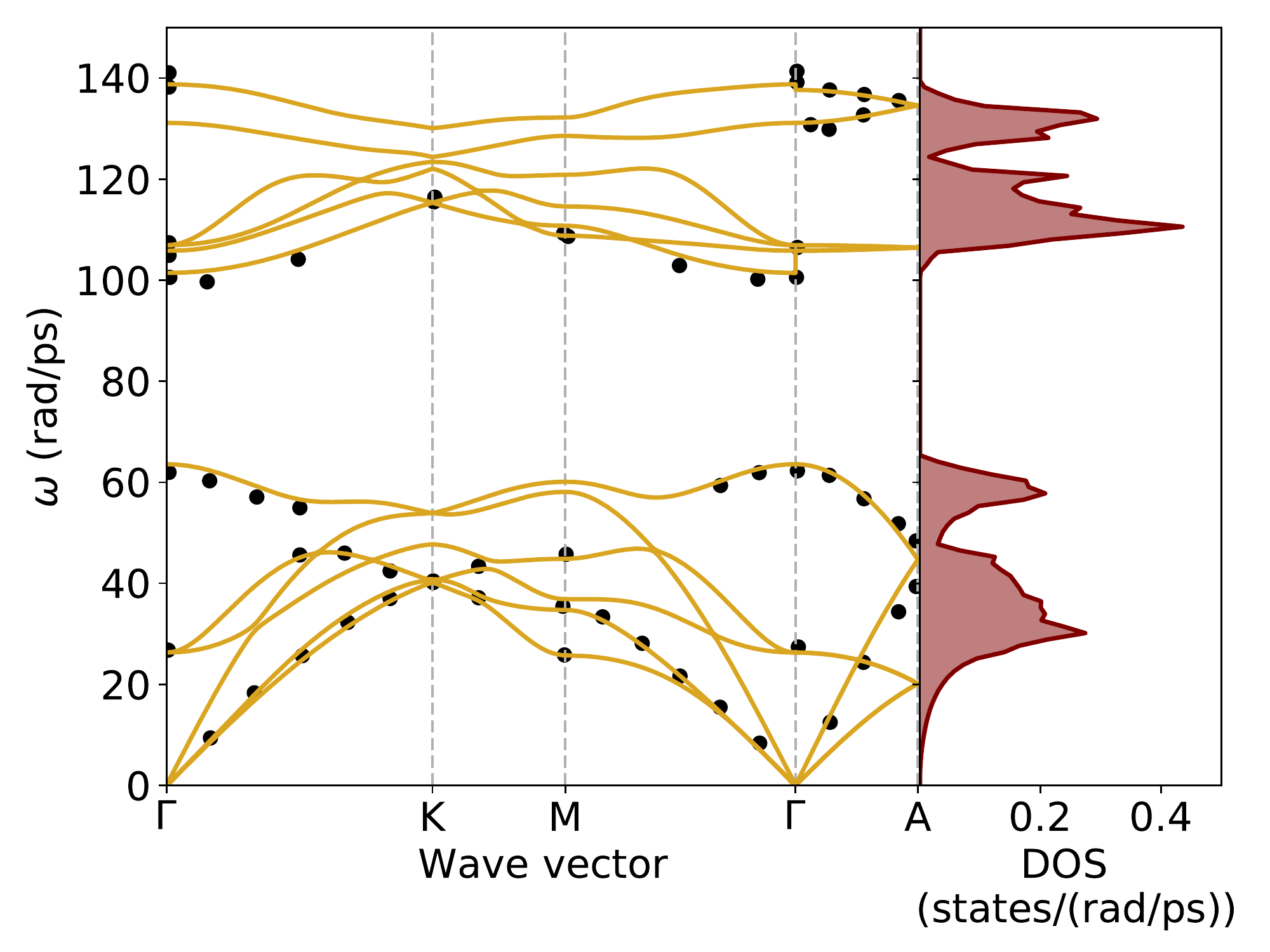} 
 \caption{Phonon dispersion and density of states for wurtzite-GaN. The calculated results are in excellent agreement with the 
 inelastic X-ray scattering results from ref.~\citenum{Ruf_PRL2001}.}
 \label{fig:phons} 
\end{figure}

First of all, we 
have verified 
that our calculations for the isotopically pure and natural single crystals (black and brown solid curves in Fig.~\ref{fig:kappaT}) agree with the previous calculation by Lindsay and Broido \cite{Lindsay_JPCM2008} (the plot displays the isotropic component 
of the conductivity as $\kappa=\frac{1}{3} \Tr(\mathbold{\kappa})$). 
Such a high thermal conductivity denotes a very weak anharmonic phonon scattering in GaN. This stems from its large acoustic-optic phonon gap due to the  
large atomic mass difference between Ga and N ($\frac{m_{\mathrm{Ga}}}{m_{\mathrm{N}}}\sim5$) \cite{Lindsay_JPCM2008, Katre_JAP2015}. The acoustic-optic gap in GaN is plotted in Fig.~\ref{fig:phons}, which shows good quantitative agreement with the inelastic X-ray scattering experiments from ref.~\citenum{Ruf_PRL2001}. 

We now proceed to determine the thermal transport signature of the different defect types, 
focusing
on the three systems whose thermal conductivities have been measured by ref.~\citenum{Simon_APL2014}: two $n$-type O-doped and one semi-insulating Mg-O co-doped sample.  To best compare the relative effect of each defect type, we plot their scattering rates ($\tau^{-1}$) at a hypothetical concentration ($c$) of $10^{20}$~cm$^{-3}$. We consider the effect of single substitutionals and vacancies, and also that of coupled dopant-vacancy and dopant-dopant complexes, including the charged states reported as most stable by ref.~\citenum{walle_first-principles_2004, lyons_computationally_2017}. 

\begin{figure}[t]
 \centering
  \includegraphics[width=9.3cm]{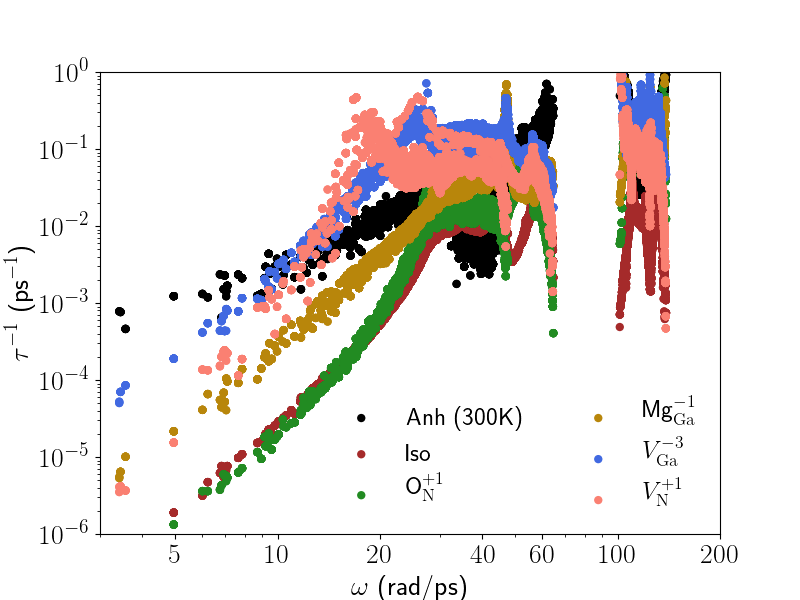} 
  \caption{Phonon scattering rates ($\tau^{-1}$) from different charged defects in GaN plotted with respect to  
  phonon frequencies ($\omega$). The defect concentration of $10^{20}$cm$^{-3}$ is considered here and the corresponding scattering rates are compared to anharmonic scattering at 300~K and isotope 
  scattering.}
  \label{fig:srates} 
\end{figure}

The case of single defects is shown in Fig.~\ref{fig:srates}. The scattering rates for gallium and nitrogen vacancies are remarkably stronger than those for the substitutional impurities. Regarding the latter, Mg$_{\mathrm{Ga}}$ scatters more strongly than either O$_{\mathrm{N}}$ or isotope impurities, both of which have similarly weak rates. 
An important consequence of this is
that any noticeable decrease in thermal conductivity upon oxygen doping 
is
most likely not 
due to direct phonon scattering by the O atoms, but rather 
by an associated change in the compensating vacancy concentration. 
Furthermore, this explains 
why the thermal conductivity is higher for the
magnesium-oxygen co-doped sample in Fig.~\ref{fig:kappaT} than 
for the oxygen-only doped
cases, even when the oxygen concentrations %
in the latter
are lower. In the case of coupled defects, the scattering rate comparison reveals another 
important
fact: the phonon scattering due to the coupled defect is very similar to the added separate contributions of the two single defects. This is shown in ESI for vacancy-oxygen and magnesium-oxygen complexes.

The above comparison of scattering rates indeed ascertains that not the oxygen, but the associated magnesium or vacancy defect will have significant impact on the thermal conductivity. However, the concentration of these associated defects, especially in the case of vacancies, is not trivial to find. For the case of magnesium-oxygen defects, only Mg$_{\mathrm{Ga}}$-O$_{\mathrm{N}}$ pair complexes form \cite{Gorczyca_PRB2000}. Simon \emph{et al.} \cite{Simon_APL2014}, measured equal concentrations of magnesium and oxygen in their co-doped sample. Using the given concentrations for the magnesium-oxygen co-doped sample in ref.~\citenum{Simon_APL2014}, we get a very good agreement of calculated thermal conductivity with the experiment upto room temperature, as seen  in Fig.~\ref{fig:kappaT}. \footnote[3]{We suspect problems in the high temperature $\kappa$ measurements from ref.~\citenum{Simon_APL2014} as they show irregularity and sudden change of $\kappa$-$T$ slope.} 
In turn,
for the case of vacancy-oxygen defects there are different opinions on whether $V_{\mathrm{Ga}}$-O$_{\mathrm{N}}$ pairs or $V_{\mathrm{Ga}}$-(O$_{\mathrm{N}}$)$_3$ clusters are favourable \cite{Simon_APL2014, Slack_JCG2002}. A few studies suggest that hydrogen is also present in 
such samples and tends to form complexes with the vacancies and oxygen defects \cite{Hauta_PRB2006, Lyons_PSSB2015}. Recent \emph{ab initio} studies support this finding based on calculated  
lower defect formation energies for $V_{\mathrm{Ga}}$-H and several $V_{\mathrm{Ga}}$-H-O$_{\mathrm{N}}$ complexes than for single isolated defects \cite{Lyons_PSSB2015, lyons_computationally_2017}. However, they do not provide any evidence of actual concentrations of the prevalent hydrogen defects and vacancies. 

Analysis of the growth process of these samples provides  
relevant information about the existing defects. Samples in ref.~\citenum{Simon_APL2014} are grown 
by the 
ammonothermal technique \cite{Dwil_JCG2009}. This growth process, performed at low temperatures and pressures, is known to produce 
samples with very low dislocation density \cite{tuomisto_vacancyhydrogen_2014, Dwil_JCG2009, Pimputkar_JCG2013}. However, higher content of point defects, mainly vacancies, is found \cite{tuomisto_vacancyhydrogen_2014,Tuomisto_JCG2010}. 
Tuomisto \emph{et al.} \cite{tuomisto_vacancyhydrogen_2014} report very high concentrations of vacancies and hydrogens in their different O-doped GaN samples and, based on their positron annihilation measurements, reveal the formation of vacancy-hydrogen complexes too \cite{tuomisto_vacancyhydrogen_2014}. Furthermore, another very interesting fact seen in their samples is that the concentrations of hydrogen, which also acts as a donor, are generally higher than those of oxygen, yet the free carrier concentration is low. This implies a high compensating vacancy content in the ammonothermally grown samples.  

\begin{figure}[t]
 \centering
  \includegraphics[width=8.3cm]{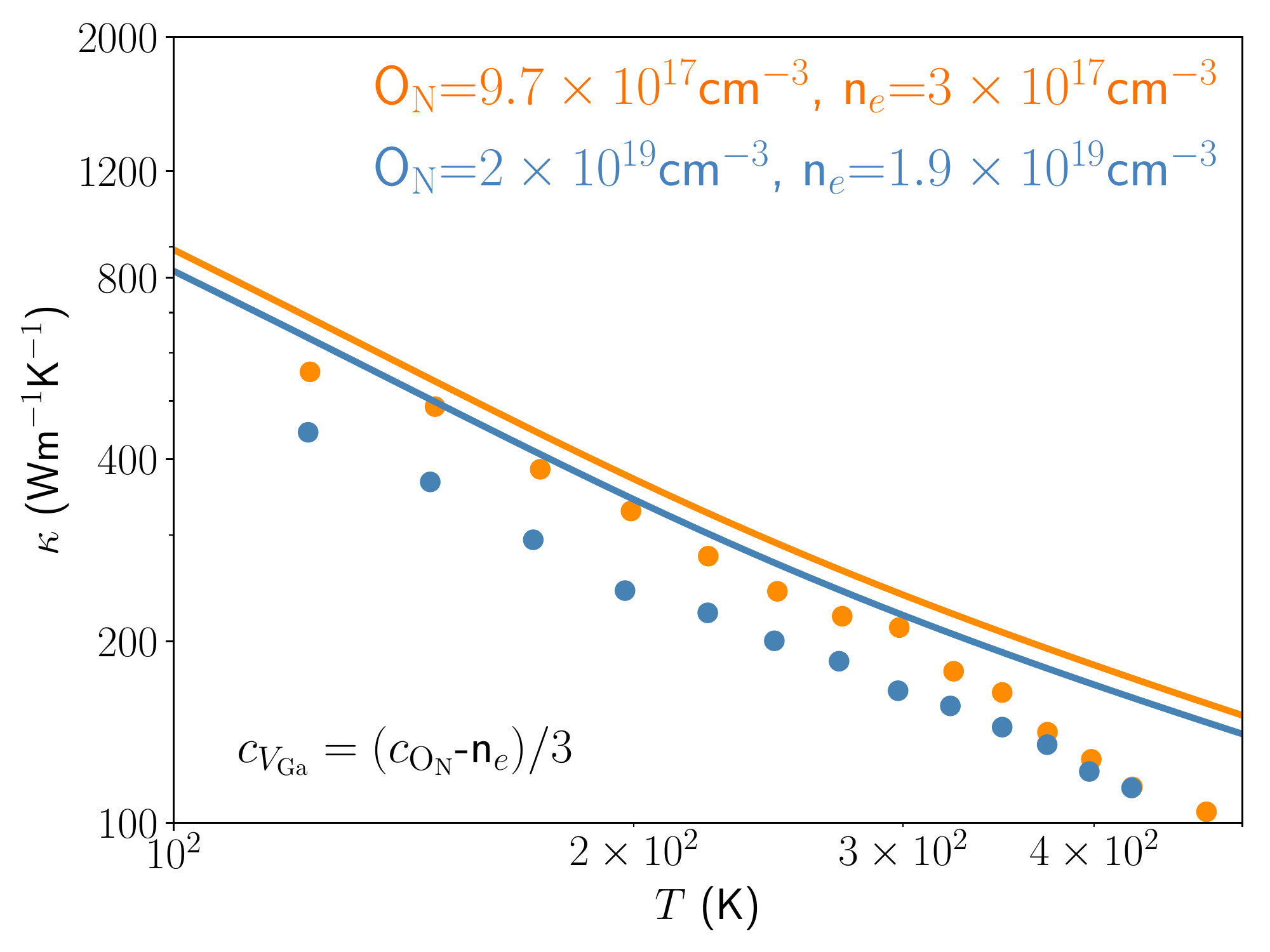} 
 \caption{GaN thermal conductivity calculated with gallium vacancies and oxygen defects only 
 conserving the free carrier concentrations in the experimental samples. The hydrogen defects are ignored in this test.  
 The predicted $\kappa$ (lines) are much higher than the experiments (circles) from ref.~\citenum{Simon_APL2014}.}
 \label{fig:kappaT_test1} 
\end{figure}

As the hydrogen concentrations are not specified for the ammonothermally grown samples in ref.~\citenum{Simon_APL2014}, we extract this information from ref.~\citenum{tuomisto_vacancyhydrogen_2014}. 
Sample-1 and sample-3 grown by Tuomisto \emph{et al.} \cite{tuomisto_vacancyhydrogen_2014} correspond to the oxygen-doped cases in ref.~\citenum{Simon_APL2014}. Thus, 
we use the hydrogen concentrations from the above mentioned two samples in ref.~\citenum{tuomisto_vacancyhydrogen_2014} and determine the vacancy concentrations by balancing the free charge carriers (n$_e$) as $c_{V_\mathrm{Ga}}=\frac{1}{3}(c_{\mathrm{O_N}}+c_{\mathrm{H}}-$ n$_e$), to calculate the thermal conductivity. We do not include the scattering contributions from hydrogen as they 
can be shown to be negligible (see ESI).
This is expected because hydrogen in GaN bonds with the nitrogen atoms surrounding vacancy \cite{lyons_computationally_2017}, 
thus acting on phonons similarly as if isotopes of N$_7^{15}$ were present.
The calculated thermal conductivity using these concentrations show remarkable agreement with the experiments for both O-doped  samples, as shown in Fig.~\ref{fig:kappaT} and Table.~\ref{tab:kl_def}. 
Thus, our calculation consistently relates the results of the two separate experiments in refs.~\citenum{tuomisto_vacancyhydrogen_2014} and \cite{Simon_APL2014}, and lends strong support to 
the hydrogen hypothesis put forth in ref.~\citenum{tuomisto_vacancyhydrogen_2014}.

\begin{table*}[t]
  \centering
  \begin{threeparttable}
  \begin{tabular}{   P{2.5cm}    P{1.5cm}  P{1.5cm}  P{1.5cm}      P{1.5cm}       P{2cm}  P{2cm} }
  \hline \hline
          Experiment          &   $ \mathrm{O_N} $  & n$_e$       &    ${\mathrm{H}}$   &   $ V_\mathrm{Ga} $ & $\kappa_{\mathrm{calc}}$ & $\kappa_{\mathrm{exp}}$  \\ 
    ref.~\citenum{Simon_APL2014} &   (cm$^{-3}$)       & (cm$^{-3}$) &   (cm$^{-3}$)         &    (cm$^{-3}$)        &   (Wm$^{-1}$K$^{-1}$)    &   (Wm$^{-1}$K$^{-1}$)    \\ \hline
     Sample-1   & 0.97$\times10^{18}$ & 0.3$\times10^{18}$ & 6$\times10^{18}$  &  2.2$\times10^{18}$ &   218   &  211      \\
     Sample-2    & 20$\times10^{18}$  &  19$\times10^{18}$  & 20$\times10^{18}$ &     7$\times10^{18}$ &   175   &  166      \\
        \hline \hline  
 \end{tabular}
  \end{threeparttable}
  \caption{ Calculated thermal conductivity ($\kappa_{\mathrm{calc}}$) at 300~K for different defect contents in O-doped GaN and compared to the 
  experiments ($\kappa_{\mathrm{exp}}$) from ref.~\citenum{Simon_APL2014}. 
  The results are calculated with vacancy concentrations as
  $c_{V_\mathrm{Ga}}$=$\frac{1}{3}(c_{\mathrm{O_{N}}}+c_{\mathrm{H}}-\mathrm{n}_e)$, where hydrogen concentrations   
  are obtained from ref.~\citenum{tuomisto_vacancyhydrogen_2014} for ammonothermally grown O-doped GaN.}
  \label{tab:kl_def}
\end{table*}

One can check that, if residual hydrogen is not considered, the measured thermal conductivity cannot be explained based on the presence of oxygen substitutionals and gallium vacancies only. In such case, the concentrations of compensating vacancies that balance the free carrier density, $c_{V_{\mathrm{Ga}}}$=$\frac{1}{3}$($c_{\mathrm{O}_{\mathrm{N}}}$-n$_e$), would be much smaller. The resulting calculated thermal conductivities are much too high,  in clear disagreement with experiment for both samples (see Fig.~\ref{fig:kappaT_test1}, and Table~I in ESI). 

Still, one might be tempted to argue that, instead of hydrogen, the presence of nitrogen vacancies ($V_\mathrm{N}$)\cite{lyons_computationally_2017, walle_first-principles_2004} could explain the observed thermal conductivities.
 Although, their coexistence with oxygen and gallium vacancies is found to be energetically unfavourable \cite{walle_first-principles_2004, Neugebauer_PRB1994, Gorczyca_PRB1999}, some
previous studies have claimed the presence of nitrogen vacancies in $n$-type GaN and identified them to be shallow donors 
like oxygen \cite{Tansley_PRB1992, Glaser_APL1993, Look_PRL1997}. This could hold true only if nitrogen vacancies in very low concentrations are present in the samples. A very 
high phonon scattering with nitrogen vacancies in Fig.~\ref{fig:srates} strengthens this possibility to explain the thermal conductivity 
of oxygen-doped samples from ref.~\citenum{Simon_APL2014}. 
To test for this,
we calculate the thermal conductivity of oxygen-doped GaN under different concentrations of nitrogen and gallium vacancies.
We adjust the $c_{V_{\mathrm{Ga}}}$ and $c_{V_{\mathrm{N}}}$ conserving $c_{\mathrm{O}_{\mathrm{N}}}$ and n$_e$ to deliberately fit to the thermal conductivity measurements \cite{Simon_APL2014}. However, these tests are unable to capture the correct $\kappa$-$T$ slopes (see Fig.~1 in ESI). Moreover, the required nitrogen vacancies, found in the order of $10^{18}$~cm$^{-3}$, are either similar to or %
just
slightly smaller than the oxygen content, as seen in Table I of the ESI. 
This is in contradiction to the \emph{ab initio} studies of the defect formation energies, 
as stated earlier \cite{walle_first-principles_2004, Neugebauer_PRB1994, Gorczyca_PRB1999}, that suggest very low nitrogen vacancy concentrations in $n$-type GaN samples. 
Therefore
the obtained 
nitrogen vacancy concentrations are improbable and fail to justify the corresponding thermal conductivities.

Thus none of the
combinations of vacancies or other impurities 
in 
the
above tests can explain the thermal conductivities in O-doped GaN samples 
while simultaneously
maintaining the free carrier density and oxygen concentrations. 
Only the thermal conductivity calculated under the assumption of residual hydrogen, with the vacancy concentrations specified by Tuomisto \emph{et al.} \cite{tuomisto_vacancyhydrogen_2014}, yields a good agreement with experimental measurements on ammonothermal samples, as seen in Fig.~\ref{fig:kappaT}.
These results are extremely significant in the field of gallium nitride as they unveil the existence of large amounts of gallium vacancies in oxygen doped samples, even larger than oxygen in some cases. Furthermore, our results also confirm the hypothesis that these vacancies indeed form complexes with hydrogen along with oxygen defects in GaN.

In conclusion, we performed \emph{ab initio} calculations of 
thermal conductivities for different O-doped and Mg-O co-doped GaN samples and found very good agreement with the previous measurements. Furthermore, 
these calculations have allowed us
to clarify the long-standing debate over the content of the prevalent vacancies in GaN samples and how they are complexed with 
other defects, providing clear evidence on the presence of hydrogen in O-doped ammonothermally grown samples. Our calculations show that the thermal conductivity measurements cannot 
be explained using the phonon scattering rates of just any defect, but only the ones actually present in the samples. 
This is because different defects have different phonon scattering strengths 
and hence have unique signatures on the thermal conductivity depending on their 
concentrations. Thus, this predictive \emph{ab inito} method to calculate thermal conductivity 
unfolds itself as a practical
and reliable way to characterize and quantify defects in materials.

The authors acknowledge the support from the Air Force Office of Scientific Research, USAF under award 
no. FA9550615-1-0187 DEF, and the European Union's Horizon 2020 Research and Innovation Programme 
[grant number 645776 (ALMA)].


\end{document}